\documentclass[a4paper,12pt]{article}
\pdfoutput=1 

\usepackage{orcidlink}
\usepackage{jheppub} 

\usepackage[T1]{fontenc} 
\usepackage{float}
\usepackage{lipsum}
\usepackage{capt-of}

\usepackage{adjustbox}

\usepackage{soul}
\setstcolor{red}

\newcommand{\mue}{\ensuremath{\mu\mathchar`-e}}
\newcommand{\mupi}{\ensuremath{\mu\mathchar`-\pi\pi^0}}
\newcommand{\emu}{\ensuremath{e\mathchar`-\mu}}
\newcommand{\epi}{\ensuremath{e\mathchar`-\pi\pi^0}}

\newcommand{\fbi}{\mathrm{fb}^{-1}}
\newcommand{\MeV}{\mathrm{MeV}}
\newcommand{\GeV}{\mathrm{GeV}}
\newcommand{\CM}{c.m.\ }
\newcommand{\effsig}{\epsilon_\mathrm{sig}}
\newcommand{\cosmiss}{\cos\theta_{\mathrm{miss}}^*}
\newcommand{\Eecl}{E_\mathrm{ECL}}
\newcommand{\Evis}{E^*_\mathrm{vis}}
\newcommand{\lOne}{\ell_1}
\newcommand{\lTwo}{\ell_2}
\newcommand{\KL}{K_L^0}
\newcommand{\KS}{K_S^0}
\newcommand{\MmissSq}{M_\mathrm{miss}^{*2}}
\newcommand{\Nbkg}{N_\mathrm{exp}^\mathrm{bkg}}
\newcommand{\Nobs}{N_\mathrm{obs}}
\newcommand{\NtwoS}{N_{\Upsilon(2S)}}

\newcommand{\Nmc}{N_\mathrm{MC}}
\newcommand{\Nsbdt}{N_\mathrm{data}^\mathrm{SB}}
\newcommand{\Nsbmc}{N_\mathrm{MC}^\mathrm{SB}}

\newcommand{\nutaubar}{\overline{\nu}_\tau}
\newcommand{\plOne}{p^*_{1}}
\newcommand{\plTwo}{p^*_{2}}
\newcommand{\ppip}{p^*_{\pi^+}}
\newcommand{\ppiz}{p^*_{\pi^0}}

\newcommand{\YoneS}{\Upsilon(1S)}
\newcommand{\YtwoS}{\Upsilon(2S)}

\newcommand{\Ytoetau}{\YtwoS \to e^{\mp}\tau^{\pm}}
\newcommand{\Ytomutau}{\YtwoS \to \mu^{\mp}\tau^{\pm}}

\newcommand{\cosAngle}{\cos\theta_{12}^*}
\newcommand{\OFastBDT}{\mathcal{O}_{\mathrm{FBDT}}}
\newcommand{\EvisHat}{\hat{E}^*_\mathrm{vis}}
\newcommand{\EmissHat}{\hat{E}^*_\mathrm{miss}}

\preprint{\vbox{\hbox{Belle Preprint 2023-14} \hbox{KEK Preprint 2023-19}}}

\title{\boldmath Search for charged-lepton flavor violation in $\Upsilon(2S) \to \ell^\mp\tau^\pm$ ($\ell=e,\mu$) decays at Belle}

\collaboration{The Belle Collaboration}
  \author{R.~Dhamija\,\orcidlink{0000-0001-7052-3163},} 
  \author{S.~Nishida\,\orcidlink{0000-0001-6373-2346},} 
    \author{A.~Giri\,\orcidlink{0000-0002-8895-0128},} 
  \author{I.~Adachi\,\orcidlink{0000-0003-2287-0173},} 
  \author{H.~Aihara\,\orcidlink{0000-0002-1907-5964},} 
  \author{D.~M.~Asner\,\orcidlink{0000-0002-1586-5790},} 
  \author{T.~Aushev\,\orcidlink{0000-0002-6347-7055},} 
  \author{R.~Ayad\,\orcidlink{0000-0003-3466-9290},} 
  \author{V.~Babu\,\orcidlink{0000-0003-0419-6912},} 
  \author{S.~Bahinipati\,\orcidlink{0000-0002-3744-5332},} 
% ...

\abstract{We report a search for the charged-lepton flavor violation in $\Upsilon(2S) \to \ell^\mp\tau^\pm$ ($\ell=e,\mu$) decays using a $25~\fbi$ $\Upsilon(2S)$ sample collected by the Belle detector at the KEKB $e^{+}$$e^-$ asymmetric-energy collider. We find no evidence for a signal and set upper limits on the branching fractions ($\mathcal{B}$) at 90\% confidence level. We obtain the most stringent upper limits: $\mathcal{B}(\Ytomutau) < 0.23 \times 10^{-6}$ and $\mathcal{B}(\Ytoetau) < 1.12 \times 10^{-6}$.}

\keywords{LFV, Belle, KEKB}

\begin{document}
\maketitle
\flushbottom

\section{Introduction}
The lepton flavor conservation is one of the accidental symmetries of the Standard Model (SM). The observation of neutrino oscillation~\cite{Super-Kamiokande:1998kpq, SNO:2002tuh} manifests the lepton flavor violation in the neutral lepton sector. In the minimal extension of the SM that explains neutrino oscillation~\cite{Super-Kamiokande:1998kpq, SNO:2002tuh}, the charged-lepton flavor violating (CLFV) transitions are mediated by $W^{\pm}$ bosons and massive neutrinos. This makes CLFV decays to be heavily suppressed on the order of $m_\nu^2/m_W^2$ such as the branching fraction $\mathcal{B}(\mu \rightarrow e\gamma) \sim 10^{-54}$~\cite{Raidal:2008jk, Teixeira:2016ecr}. Various new physics models, namely leptoquark and supersymmetry, allow enhanced decay rates for CLFV transitions~\cite{Georgi:1974sy, Pati:1974yy}.
Therefore, the observation of enhanced charged-lepton flavor violation would be clear evidence for new physics. The experimental limit on the two-body CLFV quarkonium decay provides complementary constraints on the Wilson coefficients of the effective Lagrangian of new physics models~\cite{Hazard:2016fnc}.

CLEO and BaBar studied the CLFV $\Upsilon(nS)$ $(n = 1,2,3)$ decay modes~\cite{CLEO:2008lxu, BaBar:2010vxb}, and recently Belle searched for $\Upsilon(1S) \rightarrow \ell^{\mp}\ell^{\prime\pm}$ and $\Upsilon(1S) \rightarrow \gamma \ell^{\mp}\ell^{\prime\pm}$ ($\ell = e, \mu; \ell^{\prime} = e, \mu, \tau$) decay modes~\cite{Belle:2022cce}. In this paper, we present a search for the CLFV $\YtwoS \rightarrow \ell^\mp\tau^\pm$ decays using $25~\fbi$ of the data collected at the $\YtwoS$ resonance with the Belle detector~\cite{Belle:2000cnh, Belle:2012iwr} at the KEKB asymmetric-energy $e^{+}e^{-}$ collider~\cite{Akai:2001pf, Abe:2013kxa}. The current most stringent upper limits for CLFV $\YtwoS$ decays are $\mathcal{B}(\YtwoS \rightarrow \mu^\mp \tau^\pm) < 3.3 \times 10^{-6}$ and $\mathcal{B}(\YtwoS \rightarrow e^\mp \tau^\pm) < 3.2 \times 10^{-6}$ as set by BaBar~\cite{BaBar:2010vxb}.

\section{Belle experiment}

The Belle detector is a large solid-angle magnetic spectrometer that is comprised of a silicon vertex detector (SVD), a 50-layer central drift chamber (CDC), an array of aerogel threshold Cherenkov counters (ACC), a barrel-like arrangement of time-of-flight scintillation counters (TOF), and an electromagnetic calorimeter (ECL) composed of 8736 CsI(Tl) crystals, all located inside a superconducting solenoid providing a magnetic field of 1.5~T. An iron flux-return yoke placed outside the solenoid coil is instrumented with resistive plate chambers to detect $\KL$ mesons and muons (KLM). Belle has accumulated the world's largest data sample of $e^{+}e^{-}$ collision at the center-of-mass (c.m.) energy of $10.02~\GeV$, which corresponds to 158 million $\YtwoS$ decays.

We perform the background study and the optimization of selection criteria using Monte Carlo (MC) simulated events. We use the EVTGEN~\cite{Lange:2001uf} package to generate MC events, and the detector simulation is performed with GEANT-3~\cite{Brun:1119728}. The analysis is performed in the B2BII software framework~\cite{Gelb:2018agf}, which converts the Belle to Belle~II data format. We generate 5 million signal MC events of $\YtwoS \to \ell^{\mp}\tau^{\pm}$. For background study, we use MC samples of $e^+e^- \to e^+e^-$ (Bhabha)~\cite{CarloniCalame:2003yt},
$e^+e^- \to \mu^+\mu^-$~\cite{Jadach:1999vf, PhysRevD.63.113009}, $e^+e^- \to \tau^+\tau^-$~\cite{Jadach:1999vf, PhysRevD.63.113009},
$e^+e^- \to e^+e^-\mu^+\mu^-$~\cite{Berends:1986ig}, $e^+e^- \to e^+e^-e^+e^-$~\cite{Berends:1986ig},
inclusive $\YtwoS$ decays~\cite{Lange:2001uf, Montagna:2016pvv}, and $e^+e^- \to q\bar{q}$ ($q = u,d,s,c$) processes generated with an initial state radiation (ISR) photon~\cite{Lange:2001uf, Montagna:2016pvv} at the energy of the $\YtwoS$ resonance. They correspond to $25~\fbi$ of integrated luminosity except for the sample of Bhabha events, which corresponds to $2.5~\fbi$ of integrated luminosity.

\section{Reconstruction and event selection}
\label{Sec:reconstruction}

We search for decays $\YtwoS \to \lOne^{\mp}\tau^{\pm}$ with $\tau^+ \to \lTwo^+ \nu_{\lTwo} \nutaubar$ or $\tau^+ \to \pi^+ \pi^0 \nutaubar$. Hereafter, the primary non-tau lepton is referred to as $\lOne$ and the lepton from $\tau$ as $\lTwo$,
and charge conjugation is implied. Due to copious background contributions from Bhabha and $e^+e^- \to \mu^+\mu^-$ events, we do not take the combination of the same flavored $\lOne$ and $\lTwo$. These background components also have a large contribution to the channel with $\tau^+ \to \pi^+ \nutaubar$ owing to the pion misidentification, and we do not use this channel. Therefore, our search strategy includes only $\YtwoS \to \mu^{\mp} \tau^{\pm}$ with $\tau^+ \to e^+\nu_e\nutaubar$ (\mue\ mode) or $\tau^+ \to \pi^+ \pi^0 \nutaubar$ (\mupi\ mode), and $\YtwoS \to e^{\mp} \tau^{\pm}$ with $\tau^+ \to \mu^+\nu_\mu\nutaubar$ (\emu\ mode) or $\tau^+ \to \pi^+ \pi^0 \nutaubar$ (\epi\ mode).

We require the charged particles to originate from the interaction point (IP); their distances of closest approach from the IP must be within 2.0 cm in the transverse plane and within 5.0 cm along the beam direction. Charged particles are identified based on the information from various sub-detectors~\cite{Nakano:2002jw}. Muon candidates~\cite{Abashian:2002bd} are identified using a likelihood ratio $\mathcal{R}_\mu = \mathcal{L}_{\mu}$/($\mathcal{L}_{\mu} + \mathcal{L}_{\pi} + \mathcal{L}_{K}$), where $\mathcal{L}_\mu$, $\mathcal{L}_\pi$, and $\mathcal{L}_K$ are the likelihoods for $\mu$, $\pi$, and $K$ based on the information from the KLM. All charged tracks satisfying $\mathcal{R}_\mu > 0.8$ are identified as muons with an efficiency of above 70\% and a misidentification rate of less than 3\%.  An analogous likelihood ratio $\mathcal{R}_e$ is defined for electrons~\cite{Hanagaki:2001fz}, based on the specific ionization from the CDC, the ratio of the energy deposited in the ECL to the momentum measured by the CDC and SVD combined, the shower shape in the ECL, hit information from the ACC, and matching between the position of the charged track and the ECL cluster. We require $\mathcal{R}_e > 0.8$ to distinguish electrons from charged hadrons with an efficiency of above 90\% and a misidentification rate of less than 3\%.
The energy loss of an electron via bremsstrahlung is recovered by adding back the energy of each photon found within 50 mrad of the direction of the electron track into the latter momentum.
For charged pions, we use a binary likelihood ratio $\mathcal{R}(\pi|K) = \mathcal{L}_{\pi}/(\mathcal{L}_{\pi} + \mathcal{L}_{K})$ where $\mathcal{L}_\pi$ and $\mathcal{L}_K$ are the likelihoods for $\pi$ and $K$, respectively determined from the specific ionization in the CDC, information from the TOF, and response of the ACC. We require $\mathcal{R}(\pi|K) > 0.6$ to identify charged pions that are above 87\% efficient with a kaon misidentification rate of less than 5\%.

Neutral pions ($\pi^0$) are reconstructed from photon pairs detected as ECL clusters
without associated charged particles. The energy of each photon is required to be: greater than $50~\MeV$
if detected in the barrel ($32.2^{\circ} < \theta < 128.7^\circ$), greater than $100~\MeV$ if detected in the forward endcap ($12.4^{\circ} < \theta < 31.4^\circ$), and greater than $150~\MeV$ if detected in the backward endcap ($130.7^{\circ} < \theta < 155.1^\circ$), where $\theta$ is the polar angle with respect to the direction opposite to the $e^+$ beam. The invariant mass of each photon pair is required to be within $125$ and $145~\MeV/c^2$, corresponding to three standard deviations ($\pm3\sigma$) in the $\pi^0$ mass resolution~\cite{Workman:2022ynf}.

As signal events have only two charged particles, they do not always pass the online trigger condition. The effect of finite trigger rate is studied with a trigger simulation, and only events passing the trigger simulation are chosen for simulated samples. Events of $\mue$, $\mupi$, and $\emu$ modes are mostly selected online by a trigger with two tracks and a KLM hit; in addition, ECL energy and cluster triggers contribute to the $\mue$ and $\mupi$ modes. The $\epi$ mode relies only on the ECL cluster trigger. In this analysis, we use a data sample with the following pre-selections for $\tau$-pair events: $2 \le N_{\mathrm{trk}} \le 8$, where $N_{\mathrm{trk}}$ is the number of good charged tracks; the sum of the charge of the tracks is between $-2$ and $2$; the maximum transverse momentum of the tracks is greater than $0.5~\GeV/c$; the sum of ECL energy associated with tracks is less than $5.3~\GeV$ when $2 \le N_{\mathrm{trk}} \le 4$ and the number of tracks in the barrel is less than 2~\cite{Belle:2020lfn}. These pre-selections are applied to the MC samples too.

The signature of $\YtwoS \to \lOne^{\mp}\tau^{\pm}$ signal is a high-momentum lepton $\lOne$ of momentum around $4.85~\GeV/c$ in the \CM frame. We take the highest momentum lepton in an event as $\lOne$. For the \mue\ and \emu\ modes, we consider a secondary lepton $\lTwo$ with momentum in the \CM frame ($\plTwo$) to be greater than $0.5~\GeV/c$. For the \mupi\ and \epi\ modes, we combine a $\pi^+$ with $0.3 < \ppip < 4.0~\GeV/c$ and a $\pi^0$ with $\ppiz > 0.4~\GeV/c$, where $\ppip$ and $\ppiz$ are the $\pi^+$ and $\pi^0$ momenta in the \CM frame, and require that their invariant masses lie between $0.5$ and $1.0~\GeV/c^2$. In the search for $\Ytoetau$, we require the total visible energy of the event in the \CM frame ($\Evis$), which is the sum of the energy of all neutral clusters and charged tracks, to be less than $9.8~\GeV$ in order to suppress the background from Bhabha events.

\begin{figure}[htbp]
 \begin{center}
  \includegraphics[width=0.48\linewidth]{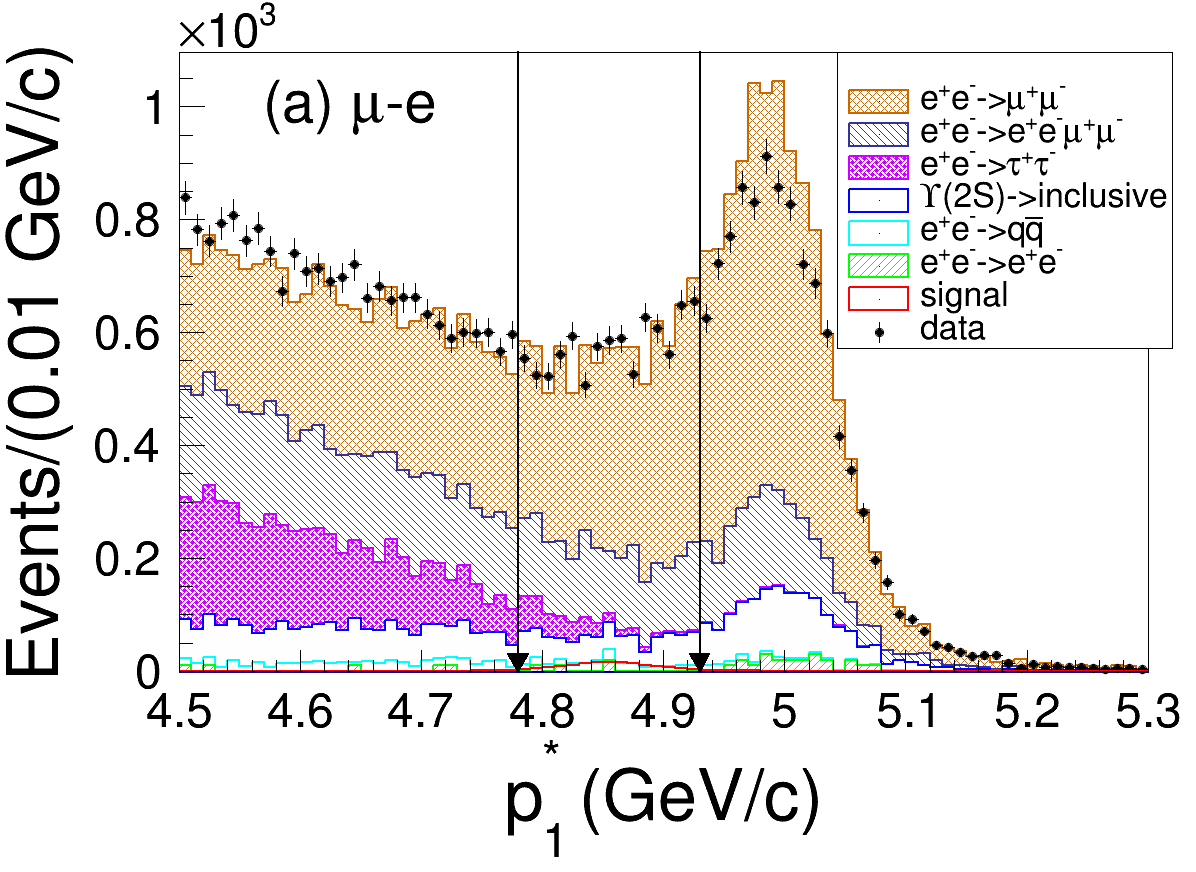}
  \includegraphics[width=0.48\linewidth]{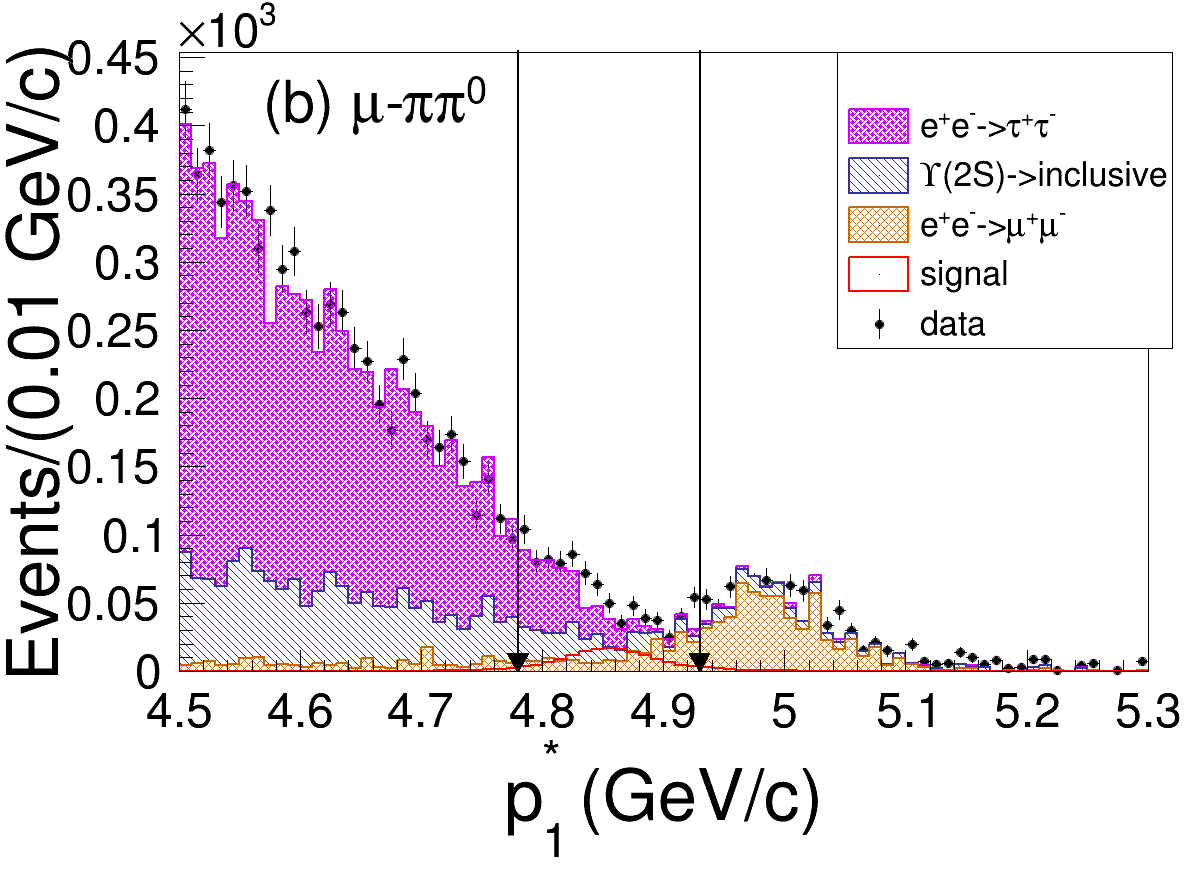}
  \includegraphics[width=0.48\linewidth]{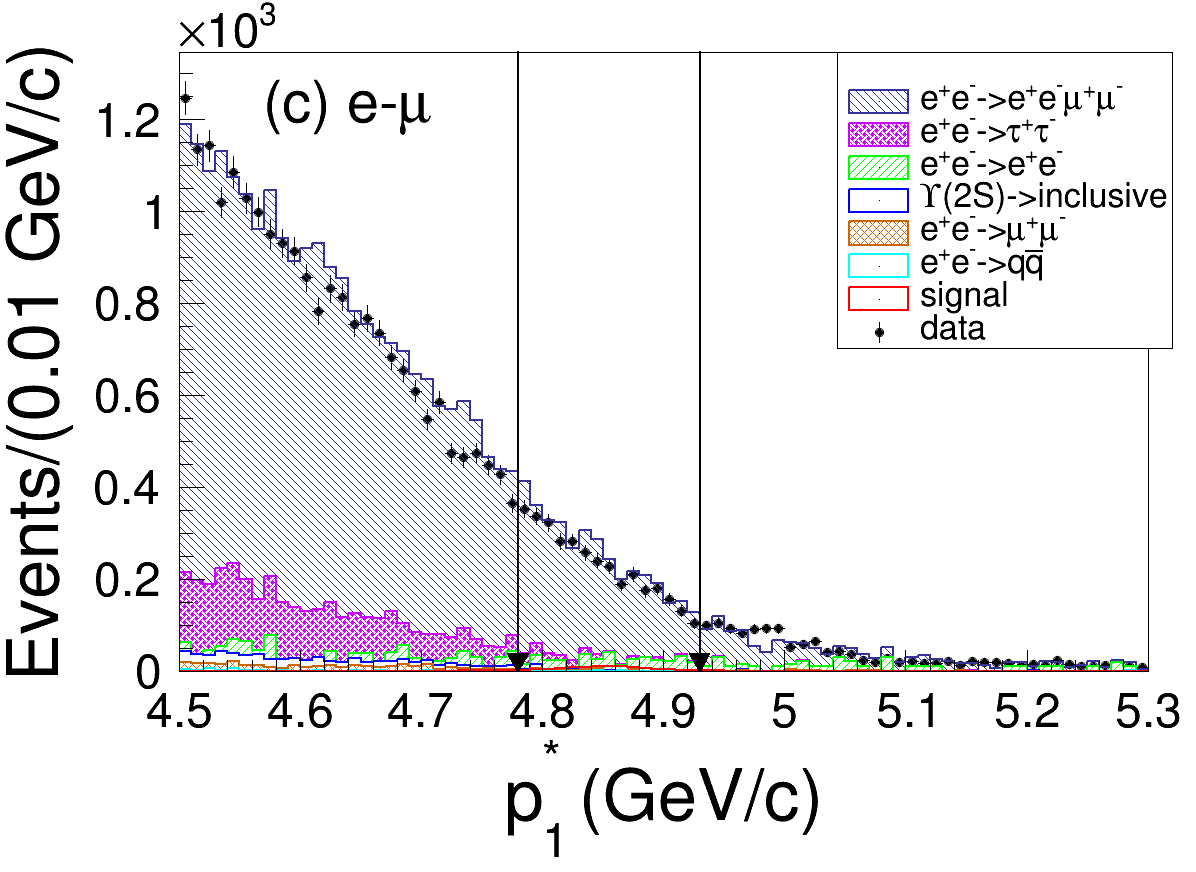}
  \includegraphics[width=0.48\linewidth]{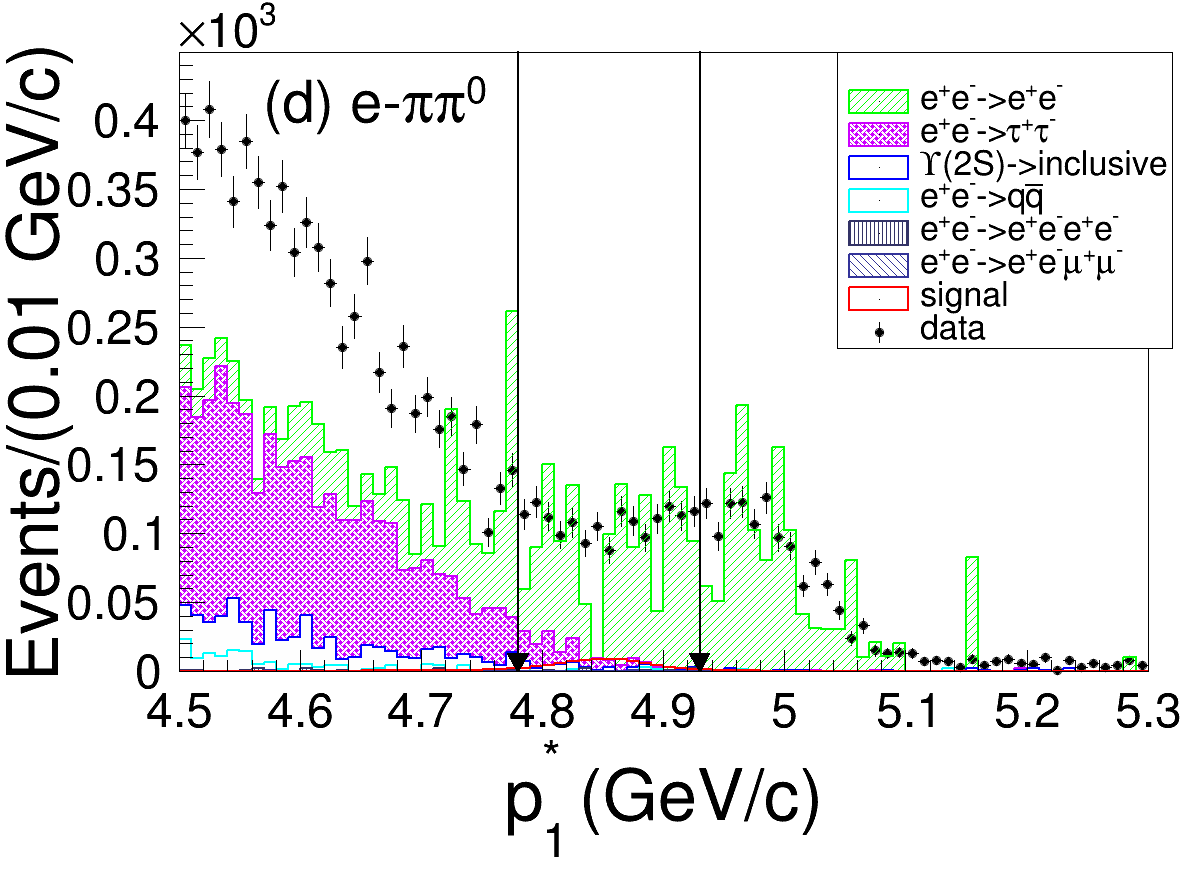}
  \caption{Distributions of $\plOne$ without the FastBDT output requirement for (a) \mue, (b) \mupi, (c) \emu, and (d) \epi.
  The downward arrows represent the signal window and we assume a high signal branching fraction $\mathcal{B} = 1 \times 10^{-5}$ to make that component visible.}
  \label{fig:pl1}
 \end{center}
\end{figure}

Figure~\ref{fig:pl1} shows the $\plOne$ distributions, where $\plOne$ is the momentum of $\lOne$ in the \CM frame
with the selections mentioned above. The signal region is chosen to be $4.78 < \plOne < 4.93~\GeV/c$, which corresponds to $2\sigma$ around the expected $\plOne$ that peaks at $4.85~\GeV/c$. The signal reconstruction efficiencies after the aforementioned selection criteria are $8.6\%$ (\mue), $8.3\%$ (\mupi), $5.9\%$ (\emu), and $4.9\%$ (\epi). After the initial selections, the dominant background contribution comes from the Bhabha, $e^{+}e^{-} \to \mu^{+}\mu^{-}$, $e^{+}e^{-} \to \tau^{+}\tau^{-}$, and $e^+e^- \to e^+e^-\mu^+\mu^-$ samples. The distribution for \epi\  has a large discrepancy between the data and the MC sample, and there exists a small discrepancy in \mue\ and \emu\ modes as well. These discrepancies may be due to an imperfect simulation of Bhabha events since these modes have a potential contribution from Bhabha events.

\begin{figure}[htbp]
 \begin{center}
  \includegraphics[width=0.48\linewidth]{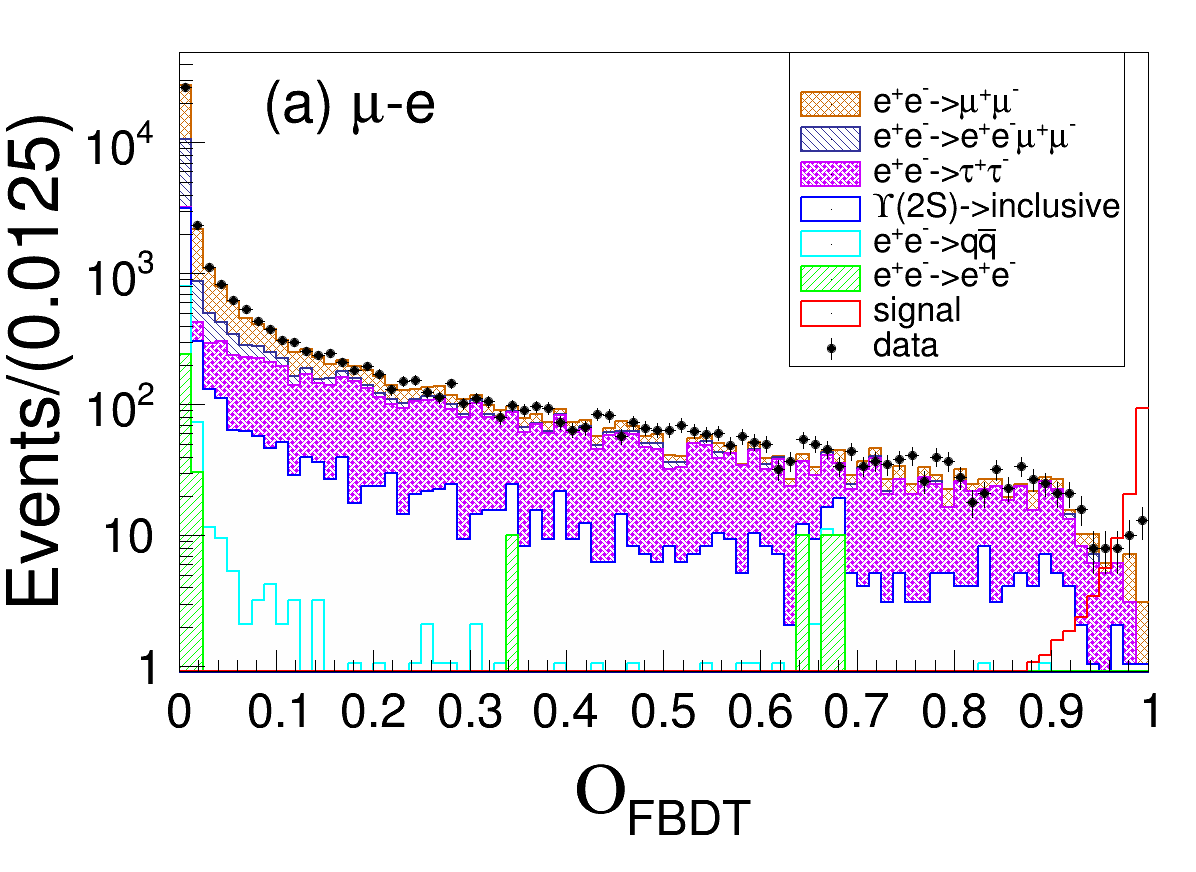}
  \includegraphics[width=0.48\linewidth]{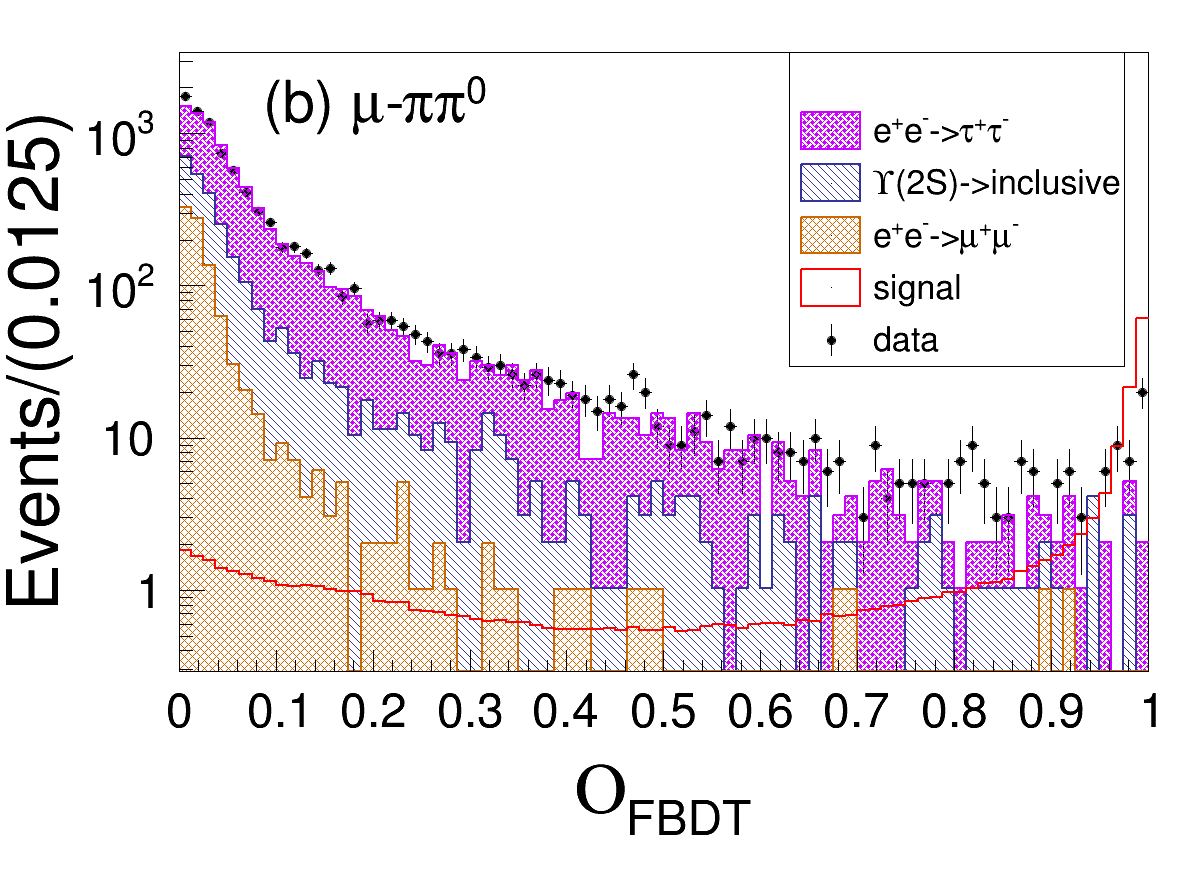}
  \includegraphics[width=0.48\linewidth]{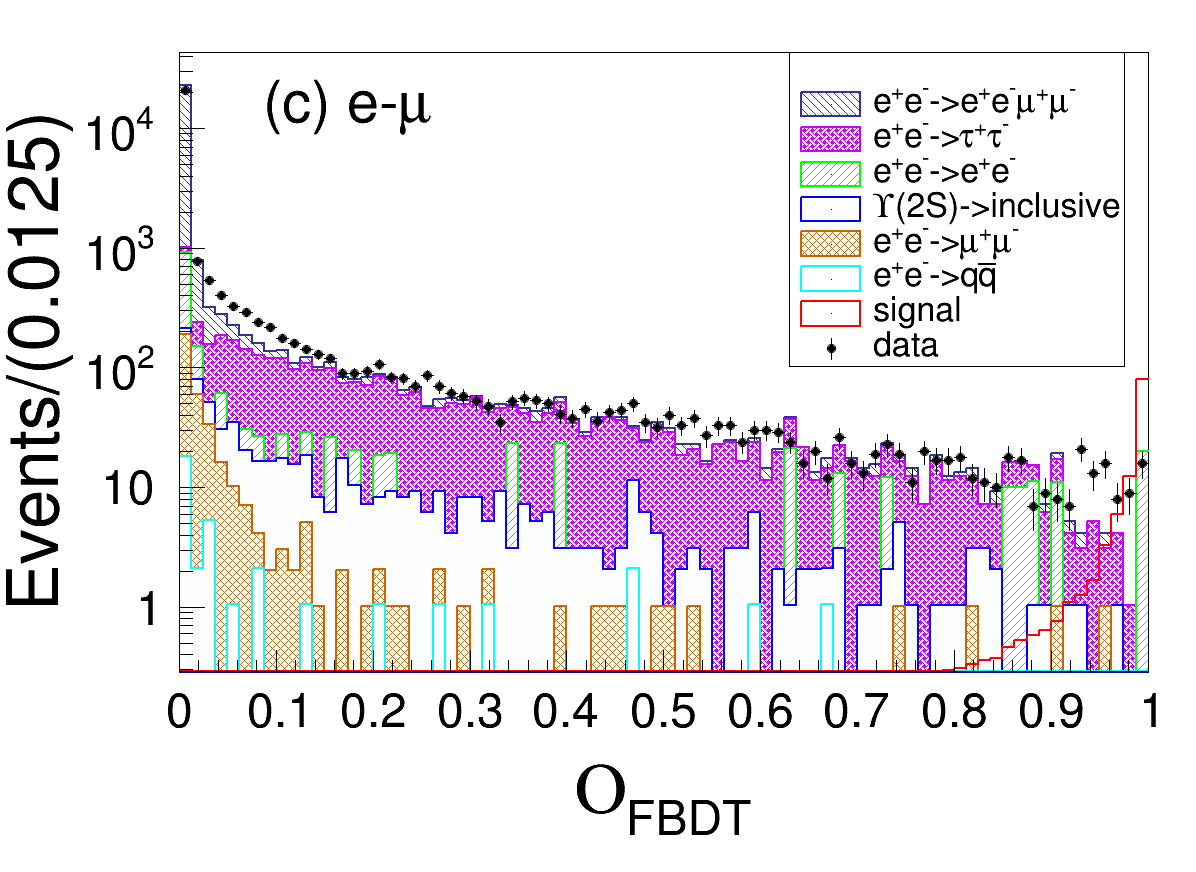}
  \includegraphics[width=0.48\linewidth]{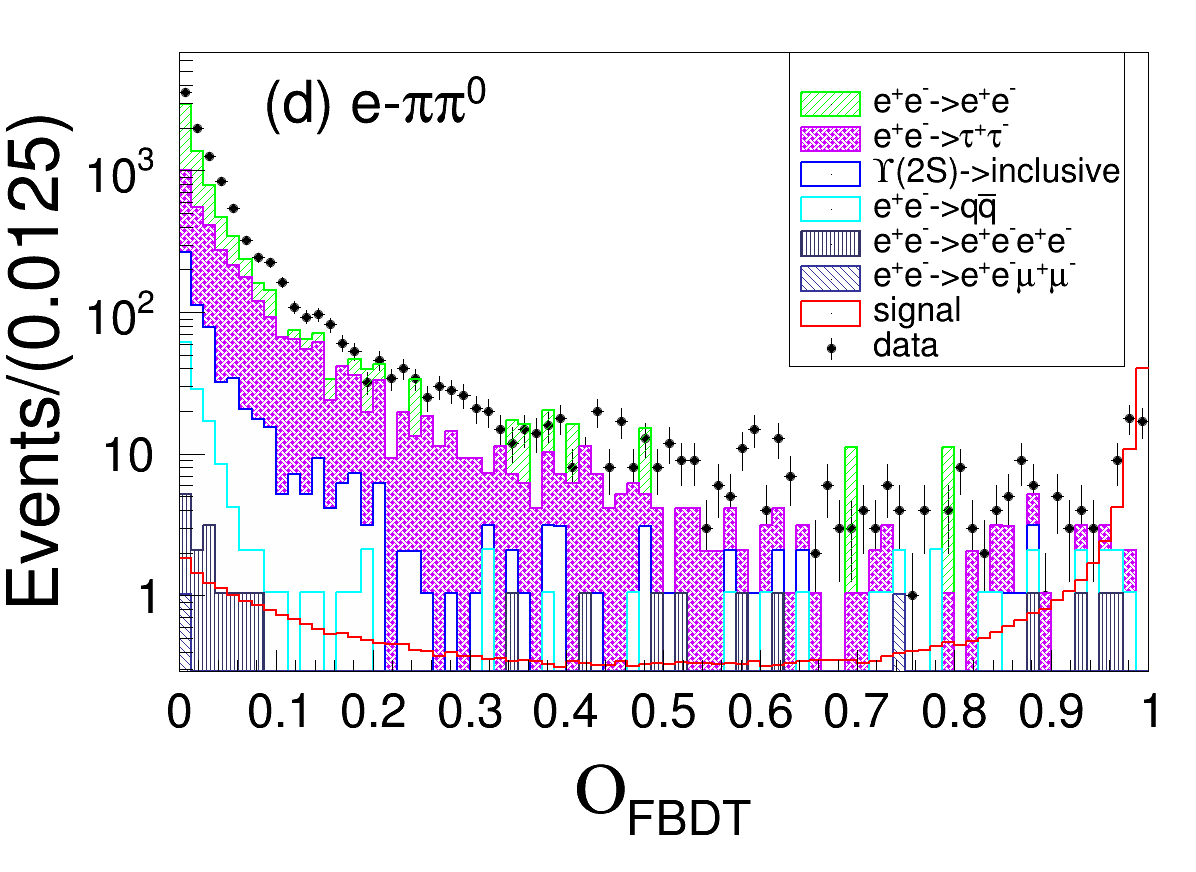}
  \caption{Distributions of $\OFastBDT$ for (a) \mue, (b) \mupi, (c) \emu, and (d) $\epi$ on a semi-logarithmic scale. The signal component assumes $\mathcal{B} = 1 \times 10^{-5}$.}
  \label{fig:FBDT}
 \end{center}
\end{figure}

A multivariate analysis (MVA) is performed to suppress the background further. We form a FastBDT~\cite{Keck:2017gsv} classifier trained with simulated samples using $\Eecl$, $\Evis$, $\MmissSq$, $\cosAngle$, and $\cosmiss$ as input discriminating variables. Here, $\Eecl$ is the sum of the energy of neutral ECL clusters that are related to the particles in the rest of the event in the lab frame, $\MmissSq$ is the invariant-mass squared of the missing momentum in the \CM frame, $\cosAngle$ is the cosine of the angle between $\plOne$ and $\plTwo$ for \mue\ and \emu\ modes or between $\plOne$ and $\ppip + \ppiz$ for \mupi\ and \epi\ modes in the \CM frame, and $\cosmiss$ is the cosine of the polar angle of the missing momentum in the \CM frame. We require $4.5 < \plOne < 5.3~\GeV/c$ for the MVA training.

Figure~\ref{fig:FBDT} shows the output of the FastBDT classifier $\OFastBDT$, which ranges from 0.0, for background-like events, to 1.0, for signal-like events. There exists a small discrepancy between data and background MC distributions at high $\OFastBDT$, which may be partly due to an imperfect simulation of Bhabha events. Since the background components do not peak in the signal region of $\plOne$ distribution, it does not introduce any bias in the signal yield estimated from $\plOne$ distribution.

At this stage, the average number of candidates in an event is $1.0$ for the $\mue$ and $\emu$ modes and $1.1$ for the $\mupi$ and $\epi$ modes. We perform the best candidate selection on the $\OFastBDT$ and retain the one with the highest $\OFastBDT$ value.

\begin{figure}[htbp]
 \begin{center}
  \includegraphics[width=0.48\linewidth]{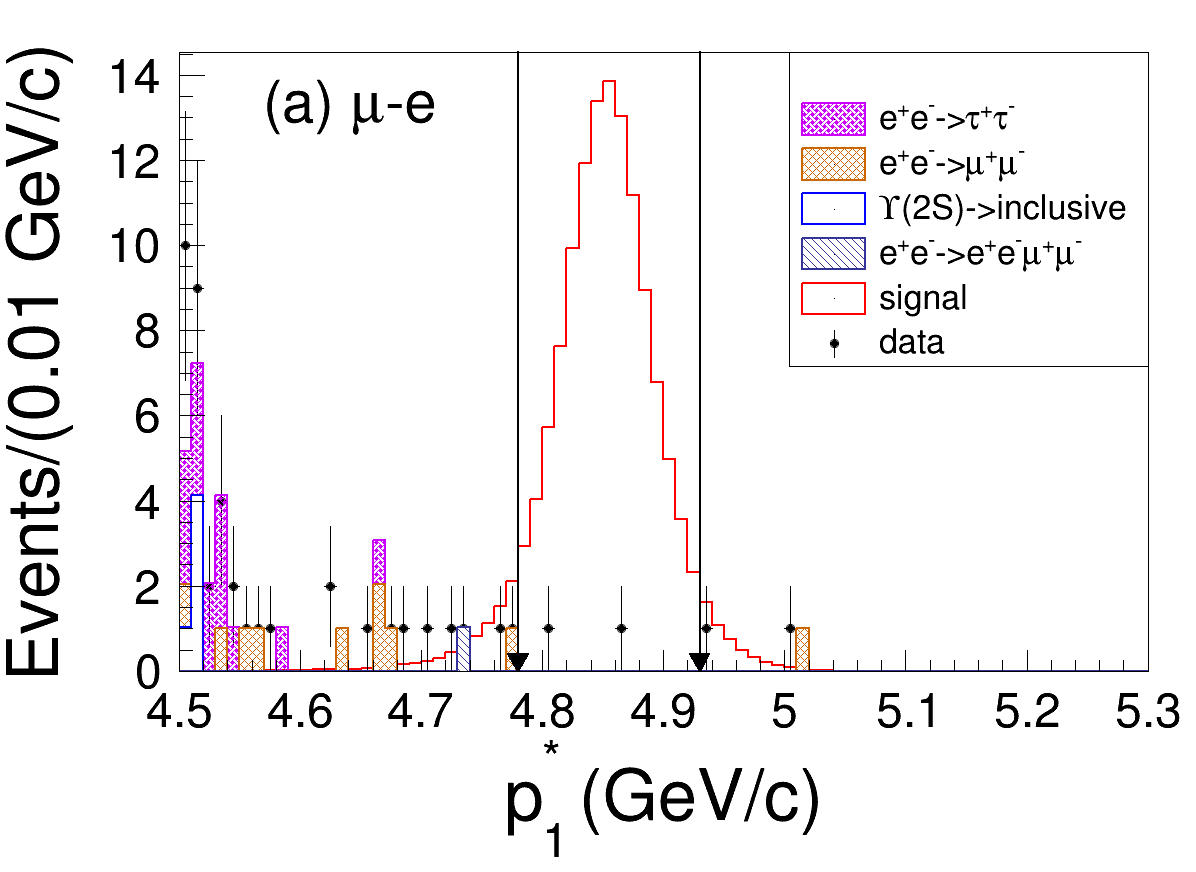}
  \includegraphics[width=0.48\linewidth]{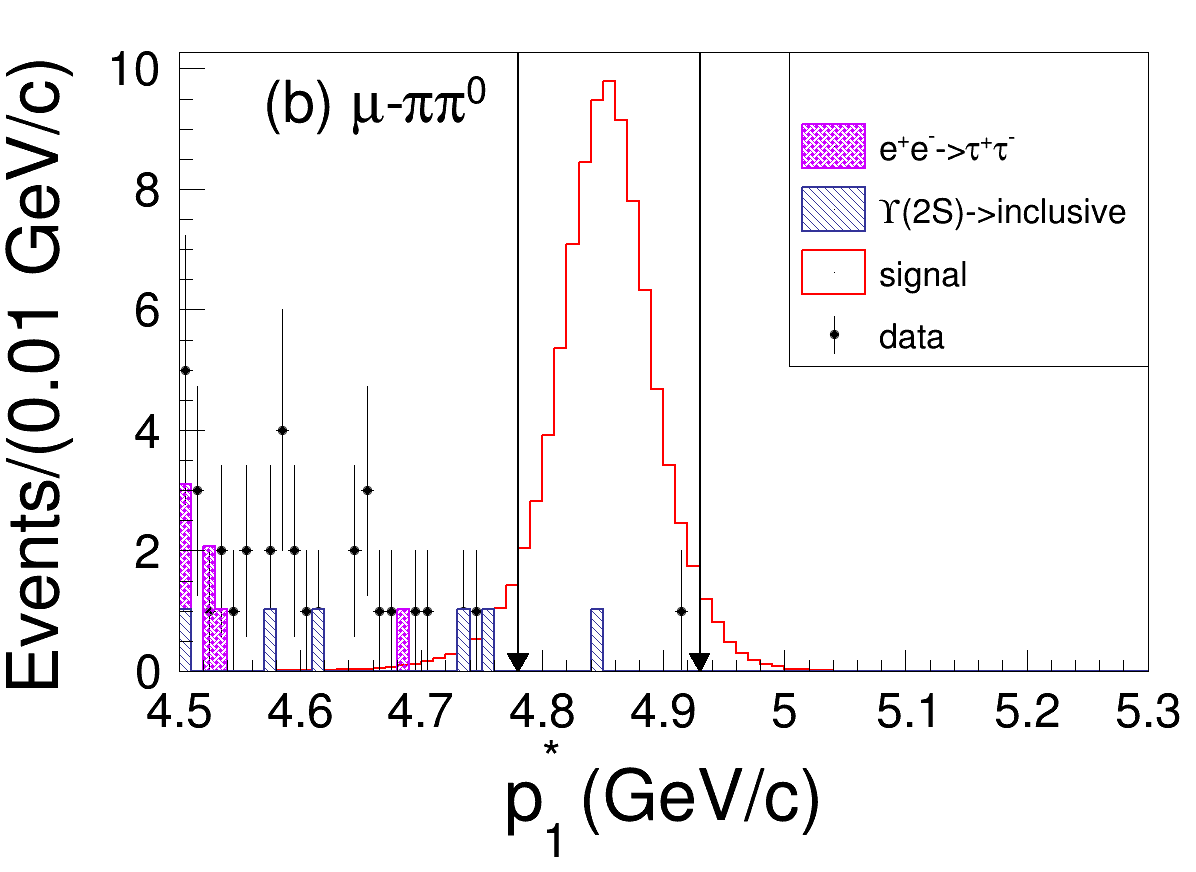}
  \includegraphics[width=0.48\linewidth]{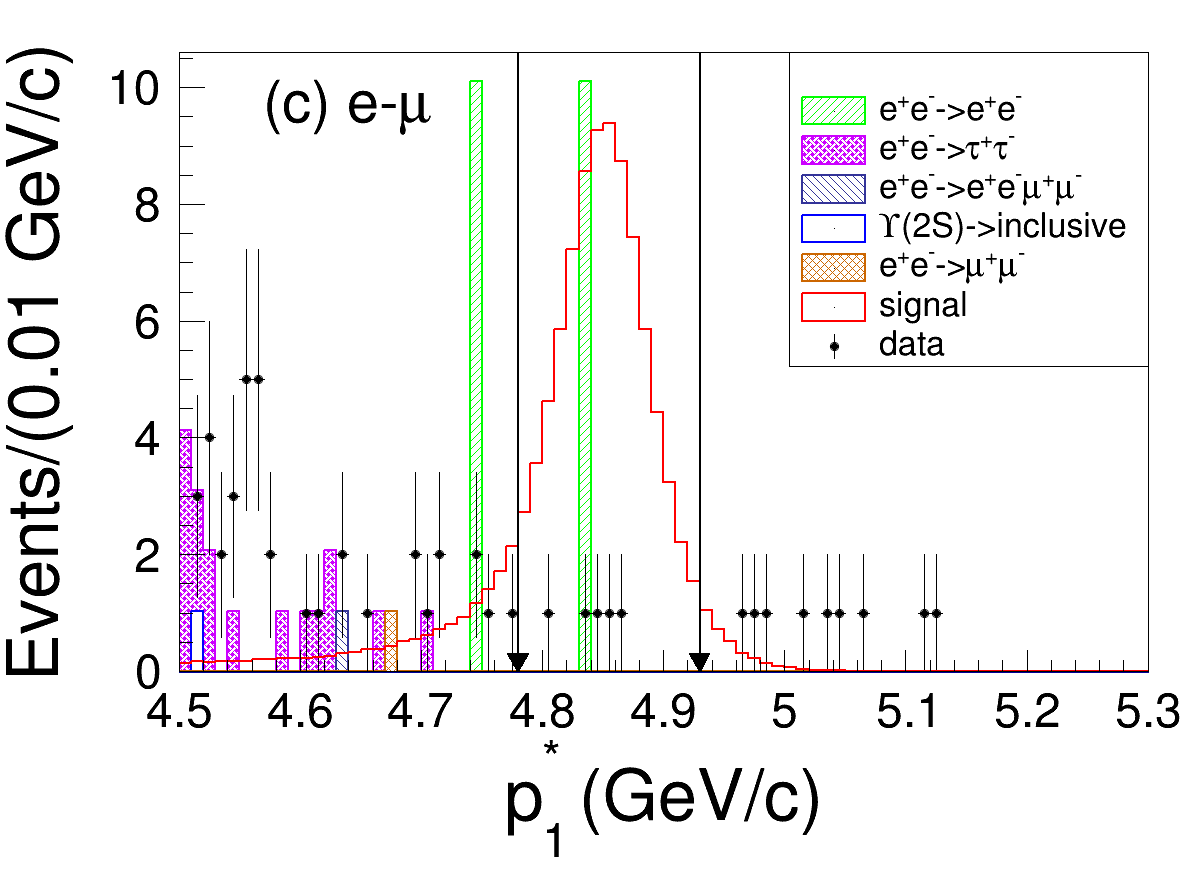}
  \includegraphics[width=0.48\linewidth]{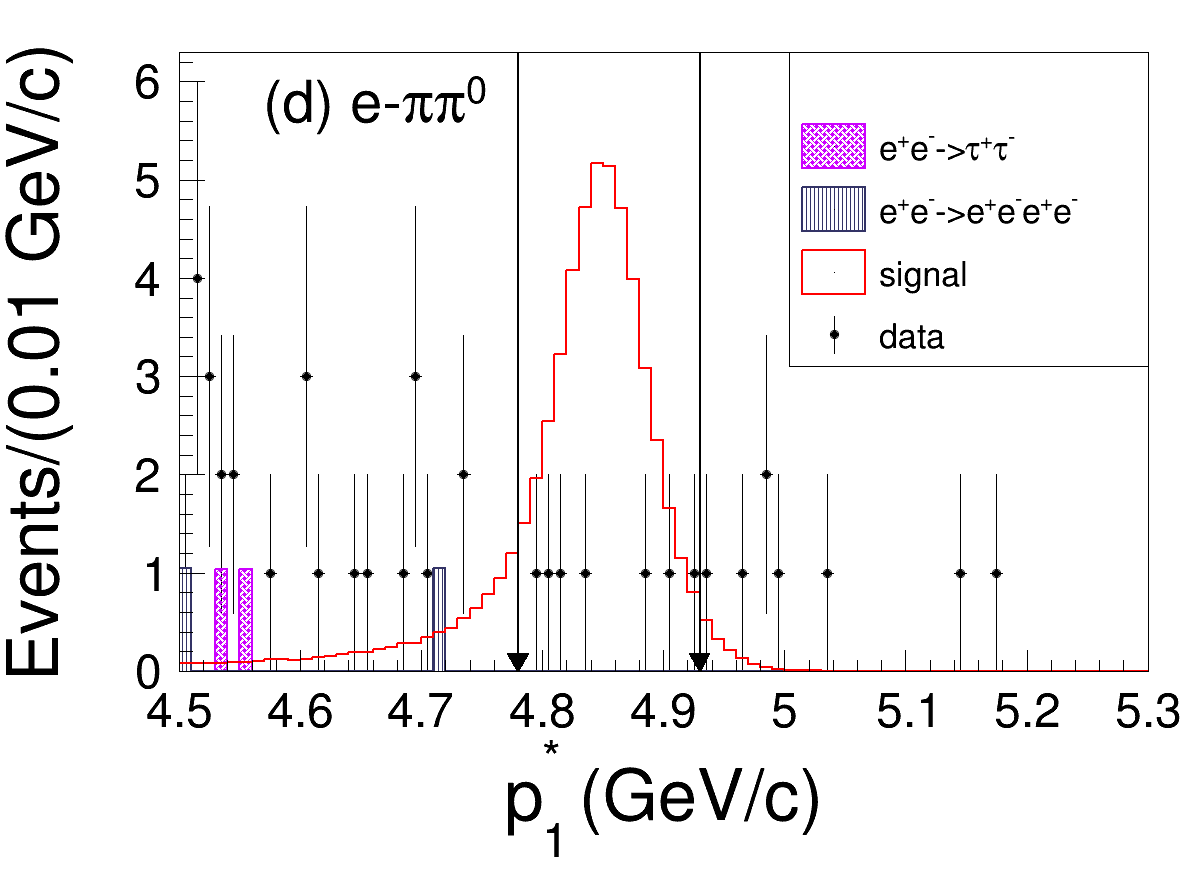}
  \caption{Distributions of $\plOne$ with the FastBDT output requirement, $\OFastBDT > 0.94$ for (a) \mue, (b) \mupi, (c) \emu, and (d) \epi. The downward arrows represent the signal window and the signal component assumes $\mathcal{B} = 1 \times 10^{-5}$.}
  \label{fig:pl1wBDT}
 \end{center}
\end{figure}

We use the figure-of-merit~\cite{Punzi:2003bu} defined as $\effsig/(3/2 + \sqrt{B})$, where $\effsig$ is the signal reconstruction efficiency and $B$ is the expected number of background events, to optimize the $\OFastBDT$ selection. The optimization is done separately for all four modes, and the optimal value depends on the mode. However, for simplicity, we choose $\OFastBDT > 0.94$ for all modes, which is slightly looser than the optimal values. This selection rejects more than $99\%$ of the background events for all the modes while retaining $86\%$, $66\%$, $89\%$, and $66\%$ of the signal events for the $\mue$, $\mupi$, $\emu$, and $\epi$ modes, respectively. The signal reconstruction efficiencies after the selection on $\OFastBDT$ are $7.4\%$, $4.8\%$, $5.3\%$, and $2.8\%$ for the $\mue$, $\mupi$, $\emu$, and $\epi$ modes, respectively. Figure~\ref{fig:pl1wBDT} shows the $\plOne$ distributions for each mode after all the selection criteria. We find three events for $\Ytomutau$ and twelve events for $\Ytoetau$ modes in the $\plOne$ signal region in the data.

\section{Systematic uncertainty}

\begin{table}[tbh]
 \begin{center}
  \caption{Summary of systematic uncertainties.}
  \vspace{2mm}
  \begin{tabular}{l|ccr}
   \hline \hline
   Source & \multicolumn{2}{c}{Systematic uncertainty (\%)} \\ \cline{2-3}
   & $\Ytomutau$  &  $\Ytoetau$ \\ \hline
   Number of $\YtwoS$ & $2.3$ & $2.3$ \\
   Tracking & $0.7$ & $0.7$ \\
   Particle identification and $\pi^0$ reconstruction & $3.4$ & $3.3$ \\
   $\tau$ branching fraction & $0.2$  & $0.2$ \\
   MVA selection & $5.1$ & $5.0$ \\
   Trigger & $2.3$ & $11.9$ \\ \hline
   Total & $7.0$ & $13.5$ \\ \hline \hline
  \end{tabular}
  \label{tab:1}
 \end{center}
\end{table}

Table~\ref{tab:1} summarizes the systematic uncertainty from various sources. The number of $\YtwoS$ ($\NtwoS$) is estimated to be $(157.8 \pm 3.6)\times 10^6$ with an uncertainty of $2.3\%$. The systematic uncertainty on the track finding efficiency is estimated to be $0.35\%$ per track from the study of partially reconstructed $D^{*+} \to D^0\pi^+$, $D^0 \to \KS\pi^+\pi^-$, $\KS \to \pi^+\pi^-$ decays. 

The systematic uncertainty due to lepton identification is studied using the control samples of $J/\psi \to \ell\ell$ and $e^+e^- \to e^+e^-\gamma^{*}\gamma^{*} \to e^+e^-\ell^+\ell^-$. The systematic uncertainty due to charged pion identification is studied using $D^{*+} \to D^{0}\pi^{+}$, $D^0 \to K^-\pi^+$ decays. The systematic uncertainty due to $\pi^0$ reconstruction is determined to be $2.3\%$ from the study on $\tau^- \to \pi^-\pi^0\nu_{\tau}$ decays. After combining these uncertainties considering the efficiency for each $\tau$ decay channel, we obtain that the uncertainty from $\Ytomutau$ ($\Ytoetau$) from particle identification and $\pi^0$ reconstruction to be $3.4\%$ ($3.3\%$). The systematic uncertainty from $\tau$ branching fraction is $0.2\%$ for both $\Ytomutau$ and $\Ytoetau$ modes.

We use the control sample $\YtwoS \to \YoneS \pi^{+}\pi^{-}$, $\YoneS \to \ell^{-}\ell^{+}$
to estimate the systematic uncertainty from MVA. We require $0.25 < M_{\pi^+\pi^-} <0.35~\GeV/c^2$, where 
$M_{\pi^+\pi^-}$ is the invariant mass of the two charged pions. We form a FastBDT classifier from the input variables $\EvisHat$, $\EmissHat$, $\Eecl$, and $\cosmiss$, where $\EvisHat$ and $\EmissHat$ are the visible energy and missing energy calculated neglecting the positively charged lepton from $\YoneS$ in order to mimic the topology of the signal $\YtwoS \to \ell^{\mp}\tau^{\pm}$ event. The FastBDT classifier is retrained using the simulated control sample applying the selection $9.3< M_{\ell^{-}\ell^{+}}<9.7~\GeV/c^2$, where $M_{\ell^-\ell^+}$ is the dilepton invariant mass. We only select the best candidate with the highest $\OFastBDT$ as mentioned earlier. We extract the signal yields from an unbinned maximum likelihood fit to $M_{\ell^-\ell^+}$ with and without the FastBDT selection, for the data and MC samples. For the data, the FastBDT selection efficiency including the best candidate is estimated to be $(11.5 \pm 0.5)\%$ and $(14.5 \pm 0.3)\%$ for the muon and electron mode, respectively, and the corresponding numbers for the MC samples are $(11.3 \pm 0.2)\%$ and $(14.0 \pm 0.3)\%$. We take the quadratic sum of the data and MC uncertainty, i.e., $5.1\%$ and $5.0\%$ respectively for $\Ytomutau$ and $\Ytoetau$ modes, as the systematic uncertainty due to MVA.

The trigger efficiency for signal events that pass all the selection criteria is estimated to be $97.6\%$ ($88.0\%$) for $\Ytomutau$ ($\Ytoetau$) from the trigger simulation. The efficiency for the electron mode is lower than for the muon mode because the former mainly relies on a single trigger condition (two tracks with a KLM hit for $\emu$ and ECL cluster trigger for $\epi$ as described in Sec.~\ref{Sec:reconstruction}). We take the difference from unity, i.e., $2.3\%$ ($11.9\%$) as the systematic uncertainty for $\Ytomutau$ ($\Ytoetau$).

Taking a quadratic sum, the total systematic uncertainty is $7.0\%$ ($13.5\%$) for $\Ytomutau$ ($\Ytoetau$) decay modes.

\section{Upper limit estimation}

We estimate the upper limits on the branching fractions with the Feldman-Cousin method, which incorporates the systematic uncertainties into this method~\cite{Feldman:1997qc, Conrad:2002kn}. We obtain the branching fractions using $\mathcal{B} = (\Nobs - \Nbkg) / (\effsig \NtwoS)$, where $\Nobs$ is the number of observed events and $\Nbkg$ is the expected number of background events. We estimate $\Nbkg$ by $\Nsbdt \times \Nmc / \Nsbmc$, where $\Nsbdt$ and $\Nsbmc$ are the numbers of data and background MC events in the sideband defined as $4.5 < \plOne < 4.7~\GeV/c$ or $5.0 < \plOne < 5.3~\GeV/c$, and $\Nmc$ is the number of background MC events in the signal region. We apply a loose selection of $\OFastBDT > 0.4$ to estimate $\Nsbmc$ and $\Nmc$, rather than the standard selection $\OFastBDT > 0.94$, and obtain the nominal value of $\Nbkg = 3.9 \pm 1.2$ for $\Ytomutau$ and $\Nbkg = 5.9 \pm 2.6$ for $\Ytoetau$. We also estimate $\Nbkg$ requiring $0.2<\OFastBDT<0.4$ for $\Nsbmc$ and $\Nmc$ and take the difference from the nominal value of $\Nbkg$, $1.28$ and $0.46$ for the $\Ytomutau$ and $\Ytoetau$ modes, respectively, as a systematic uncertainty of $\Nbkg$. As a result, we obtain $\Nbkg = 3.9 \pm 1.8$ for $\Ytomutau$ and $\Nbkg = 5.9 \pm 2.6$ for $\Ytoetau$ modes.

We find $\Nobs = 3$ for $\Ytomutau$, which is consistent with the expectation, and $\Nobs = 12$ for $\Ytoetau$. The latter is larger than the background expectation but the signal is not significant; the probability of obtaining 12 or more events with $\Nbkg = 5.9 \pm 2.6$ is 8.1\%. We calculate the upper limits on branching fractions at $90\%$ confidence level (CL) using the POLE (POissonian Limit Estimator) program~\cite{CONRAD2004117, Pole}. The estimated upper limits are $\mathcal{B}(\Ytomutau) < 0.23 \times 10^{-6}$ and $\mathcal{B}(\Ytoetau) < 1.12 \times 10^{-6}$ as listed in Table \ref{tab:2}. We obtain 14 (3) times better upper limits for $\Ytomutau$ ($\Ytoetau$) as compared to previous results from BaBar~\cite{BaBar:2010vxb}. This is the first measurement on CLFV $\YtwoS$ decays from Belle.

\begin{table}[htbp]
 \begin{center}
  \caption{Final results for $\Ytomutau$ and $\Ytoetau$ modes.}
 \vspace{2mm}
 \begin{tabular}{c|c|c|c|c}
   \hline\hline
   Modes & $\effsig$ ($\%$) & $\Nbkg$ & $\Nobs$ & $\mathcal{B}$ @ 90\% CL\\ 
   \hline
   $\Ytomutau$ & $12.3 \pm 0.8$ & $3.9 \pm 1.8$ & 3 & $<0.23 \times 10^{-6}$\\
   $\Ytoetau$ & $8.1 \pm 1.1$ & $5.9 \pm 2.6$ & 12 & $<1.12 \times 10^{-6}$ \\ 
   \hline \hline
  \end{tabular}
  \label{tab:2}
 \end{center}
\end{table}

\section{Summary}
In this paper, we have conducted a search for charged-lepton flavor violation in $\YtwoS \to \ell^\mp \tau^\pm$ decays using $25~\fbi$ data collected at the $\YtwoS$ resonance with the Belle detector. Without any evidence for such a signal, we set the upper limits on the branching fractions at 90\% CL on $\YtwoS$ decays. Table \ref{tab:2} summarizes the results obtained for these decays. Our upper limits are more stringent than the previous world's best results from the BaBar collaboration~\cite{BaBar:2010vxb}.

\acknowledgments

This work, based on data collected using the Belle detector, which was
operated until June 2010, was supported by
the Ministry of Education, Culture, Sports, Science, and
Technology (MEXT) of Japan, the Japan Society for the
Promotion of Science (JSPS), and the Tau-Lepton Physics
Research Center of Nagoya University;
the Australian Research Council including grants
DP180102629, 
DP170102389, 
DP170102204, 
DE220100462, 
DP150103061, 
FT130100303; 
Austrian Federal Ministry of Education, Science and Research (FWF) and
FWF Austrian Science Fund No.~P~31361-N36;
the National Natural Science Foundation of China under Contracts
No.~11675166,  
No.~11705209;  
No.~11975076;  
No.~12135005;  
No.~12175041;  
No.~12161141008; 
Key Research Program of Frontier Sciences, Chinese Academy of Sciences (CAS), Grant No.~QYZDJ-SSW-SLH011; 
Project ZR2022JQ02 supported by Shandong Provincial Natural Science Foundation;
the Ministry of Education, Youth and Sports of the Czech
Republic under Contract No.~LTT17020;
the Czech Science Foundation Grant No. 22-18469S;
Horizon 2020 ERC Advanced Grant No.~884719 and ERC Starting Grant No.~947006 ``InterLeptons'' (European Union);
the Carl Zeiss Foundation, the Deutsche Forschungsgemeinschaft, the
Excellence Cluster Universe, and the VolkswagenStiftung;
the Department of Atomic Energy (Project Identification No. RTI 4002) and the Department of Science and Technology of India;
the Istituto Nazionale di Fisica Nucleare of Italy;
National Research Foundation (NRF) of Korea Grant
Nos.~2016R1\-D1A1B\-02012900, 2018R1\-A2B\-3003643,
2018R1\-A6A1A\-06024970, RS\-2022\-00197659,
2019R1\-I1A3A\-01058933, 2021R1\-A6A1A\-03043957,
2021R1\-F1A\-1060423, 2021R1\-F1A\-1064008, 2022R1\-A2C\-1003993;
Radiation Science Research Institute, Foreign Large-size Research Facility Application Supporting project, the Global Science Experimental Data Hub Center of the Korea Institute of Science and Technology Information and KREONET/GLORIAD;
the Polish Ministry of Science and Higher Education and
the National Science Center;
the Ministry of Science and Higher Education of the Russian Federation, Agreement 14.W03.31.0026, 
and the HSE University Basic Research Program, Moscow; 
University of Tabuk research grants
S-1440-0321, S-0256-1438, and S-0280-1439 (Saudi Arabia);
the Slovenian Research Agency Grant Nos. J1-9124 and P1-0135;
Ikerbasque, Basque Foundation for Science, Spain;
the Swiss National Science Foundation;
the Ministry of Education and the Ministry of Science and Technology of Taiwan;
and the United States Department of Energy and the National Science Foundation.
These acknowledgements are not to be interpreted as an endorsement of any statement made by any of our institutes, funding agencies, governments, or
their representatives.
We thank the KEKB group for the excellent operation of the
accelerator; the KEK cryogenics group for the efficient
operation of the solenoid; and the KEK computer group and the Pacific Northwest National
Laboratory (PNNL) Environmental Molecular Sciences Laboratory (EMSL)
computing group for strong computing support; and the National
Institute of Informatics, and Science Information NETwork 6 (SINET6) for
valuable network support.
S.N.\ is supported by JSPS KAKENHI grant JP17K05474, JP23K03442.

\bibliographystyle{JHEP.bst}
\bibliography{bibliography.bib}

\end{document}